\documentclass[a4paper,11pt]{article}
\usepackage{pos}

\title{ The Ultra-High-Energy Source MGRO J1908+06}
 \ShortTitle{The UHE Source MGRO J1908+06}

\author*[a]{Kelly Malone}

\affiliation[a]{Los Alamos National Laboratory,\\
  Los Alamos, NM, USA}

\forColl{HAWC} 

\emailAdd{kmalone@lanl.gov}

\abstract{The TeV gamma-ray source MGRO J1908+06 is one of the highest-energy sources known, with observed emission by the High Altitude Water Cherenkov (HAWC) Observatory extending well past 100 TeV. The source exhibits both energy-dependent morphology and a spatially-dependent spectral index. The emission is likely to be dominantly leptonic, and associated with the radio-quiet PSR J1907+0602. However, one-population models do not describe the data well; a second particle population is needed to explain the shape of the spectral energy distribution at the highest energies. This component can be well-described by either leptonic or hadronic hypotheses. We discuss this feature and implications for detection by multi-wavelength and multi-messenger experiments.}

\FullConference{37$^{\rm{th}}$ International Cosmic Ray Conference (ICRC 2021)\\
		July 12th -- 23rd, 2021\\
		Online -- Berlin, Germany}

\begin{document}
\maketitle

\section{Previous observations}

The ultra-high-energy source MGRO J1908+063 was discovered by the Milagro experiment in 2007~\cite{Milagro} and subsequently observed by H.E.S.S., ARGO, VERITAS, and HAWC, among others~\cite{HESS,ARGO,VERITAS,3hwc}.  The source is exceptionally bright, spatially extended, and has a hard spectrum with little or no curvature up to $\sim$ 20 TeV.  Recent observations by HAWC indicate that the source emits well past 100 TeV~\cite{HECatalog}. The maximum photon energy is 218 TeV at the 95$\%$ confidence level~\cite{lorentz}.

The region surrounding the TeV emission is fairly crowded, with many possible counterparts to the emission. There are two pulsars: PSR J1907+0602 and PSR J1907+0631, as well as SNR G40.5-0.5. PSR J1907+0602 is an exceptionally powerful, radio-faint pulsar, with an $\dot{E}$ of 2.8 $\times$ 10$^{36}$ erg/s~\cite{Fermi}.  Historically, the TeV emission has been attributed to the nebula surrounding this pulsar.  There are also several molecular clouds in the region~\cite{radio}.

The pulsar wind nebula has not been detected at energies below GeV; for example no extended emission has been observed in the X-ray band. Extended GeV emission was undetected until recently~\cite{FermiPWN}. An analysis using Fermi-LAT data found that there are likely two components in the GeV region: a lower-energy ($<$ 10 GeV) component associated with the SNR and a higher-energy component attributed to inverse Compton scattering.

Due to the hard spectrum of the TeV emission, this source has long been considered a potential neutrino source.  It has the best p-value for a Galactic source in IceCube catalog searches, although it is still consistent with background~\cite{IceCube}. 

In this proceeding, we investigate the spectrum and morphology of the source using HAWC data. We will then discuss possible origins of the TeV emission. Note that the results presented here will be submitted for publication soon, with the analysis and results described in much more detail. 

\section{HAWC results}

The analysis presented here uses data from the High Altitude Water Cherenkov (HAWC) Gamma-Ray Observatory, which is an extensive air shower array located at an altitude of 4100 meters in the state of Puebla, Mexico. It is optimized to detect gamma rays with energies between a few hundred GeV and a few hundred TeV. For more details about HAWC, refer to ~\cite{Smith2015}. 

This analysis uses 1343 days of data with reconstructed energies above 1 TeV. The energy is estimated using the ``ground parameter" energy estimation method, which relies on the charge measured in the PMTs 40 meters from the shower core along with the zenith angle of the event~\cite{Crab}. 

\begin{figure*}
\centering
\includegraphics[width=0.5\textwidth]{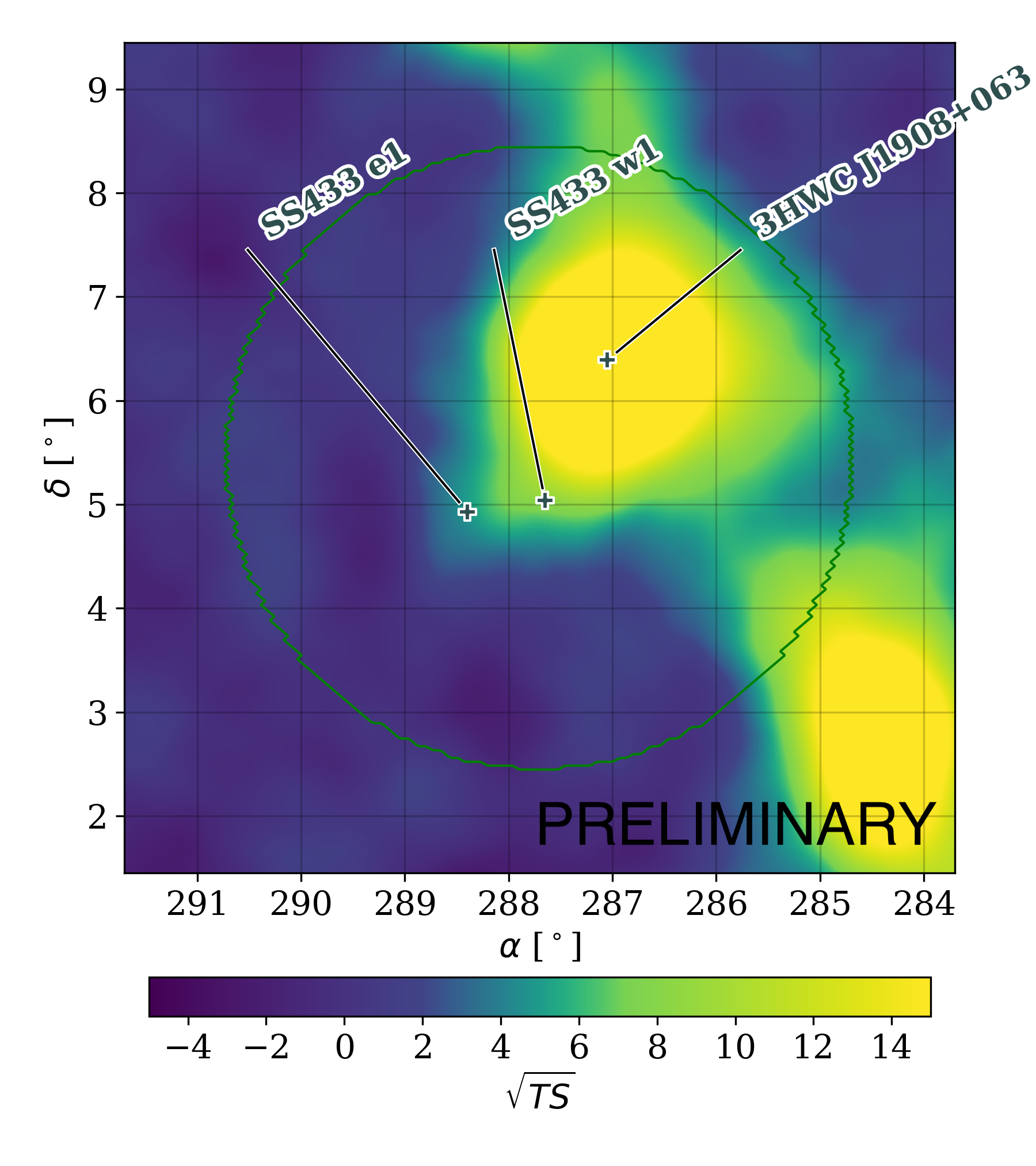}
\caption{HAWC significance map of the region, for 1 $<$ $E_{reconstructed} < 316$ TeV. The green circle denotes the region of interest for the likelihood fit. The three sources included in the model are labeled. The color scale is set to make it easy to see where the 5$\sigma$ threshold is but is saturated at the end of the scale. The maximum significance in the ROI is 38.8$\sigma$.}
\label{fig:sig}
\end{figure*}

The spectrum and morphology are simultaneously determined via a likelihood fit. There are three sources in the model: 3HWC J1908+063 (where 3HWC refers to HAWC's third catalog of sources~\cite{3hwc}, the most recent available) as well as the lobes of SS433, which are spatially coincident with the edge of the source~\cite{ss433}.  A significance map of the region can be seen in Figure ~\ref{fig:sig}. The likelihood fit is performed using the HAWC Accelerated Likelihood (HAL)\footnote{\url{https://github.com/threeML/hawc_hal}} plugin to the 3ML (the Multi-mission maximum likelihood) software package~\cite{threeml}. 

The best-fit spectrum is a log parabola shape detected in the range from $\sim$500 GeV to $\sim$200 TeV.  The best-fit morphology is a diffusion morphology where the particles are continuously injected from the center of the source.  The details of the diffusion model can be found in ~\cite{Geminga}. 

\subsection{Morphology studies}

The source exhibits energy-dependent morphology, with the size of the source decreasing with energy. This could be an indication that the emission is predominantly leptonic, as the higher-energy particles are not able to travel far before cooling. The source remains spatially extended at the highest energies. 

The spectral index can also be spatially resolved. If one draws four concentric rings around the center of the source and fits the spectrum to a power-law in each one, a statistically significant softening of the spectrum is observed as a function of distance from the center of the source.  This is also an indication of leptonic emission, as electrons far away from the center of the source are likely older and have cooled. No spectral change is expected for hadrons, since the proton cooling timescale is approximately infinitely long. 

\subsection{Potential spectrum hardening feature}

A potentially interesting feature can be seen at the highest energies. A standard HAWC analysis contains energy bins that are a quarter-decade wide in log-energy space. When the last three typical energy bins are subdivided into six smaller bins of equal size, an apparent flattening in the spectrum can be seen by eye, with the flux points deviating from the best-fit log parabola spectrum.

This effect is not presently statistically significant; the amount of the deviation from the best-fit log parabola is approximately 2$\sigma$.  However, if this feature is shown to be significant with either improved reconstruction algorithms or more data, it could be indicative of a second population of particles contributing to the gamma-ray emission at the highest energies. 

For a more in-depth discussion of this spectrum hardening feature, see ~\cite{spectralHardening} in these proceedings.

\section{The origin of the emission}

Modeling indicates that one-population models are unlikely. Looking at the gas in the region, a one-population hadronic model is difficult to explain as there is not enough energy available to account for the source's observed flux. It is difficult to fit a one-population leptonic model to the observed TeV shape.

Therefore, two populations of particles are needed to explain the shape of the spectrum in the TeV range.  One potential model consists of a leptonic component that provides an explanation for the bulk of the emission, with a second component that contributes mainly at the highest energies. HAWC's uncertainties are currently too large to determine if this second component is leptonic or hadronic in origin.

The first leptonic component is likely the PWN associated with PSR J1907+0602.  The second component could be produced by a variety of different explanations.  If it is leptonic, it could originate from the second pulsar in the region, PRS J1907+0631. If hadronic, it could be linked to the supernova remnant. More exotic explanations, such as hadron acceleration in PWN, are also plausible. 

\begin{figure*}
\centering
\includegraphics[width=0.8\textwidth]{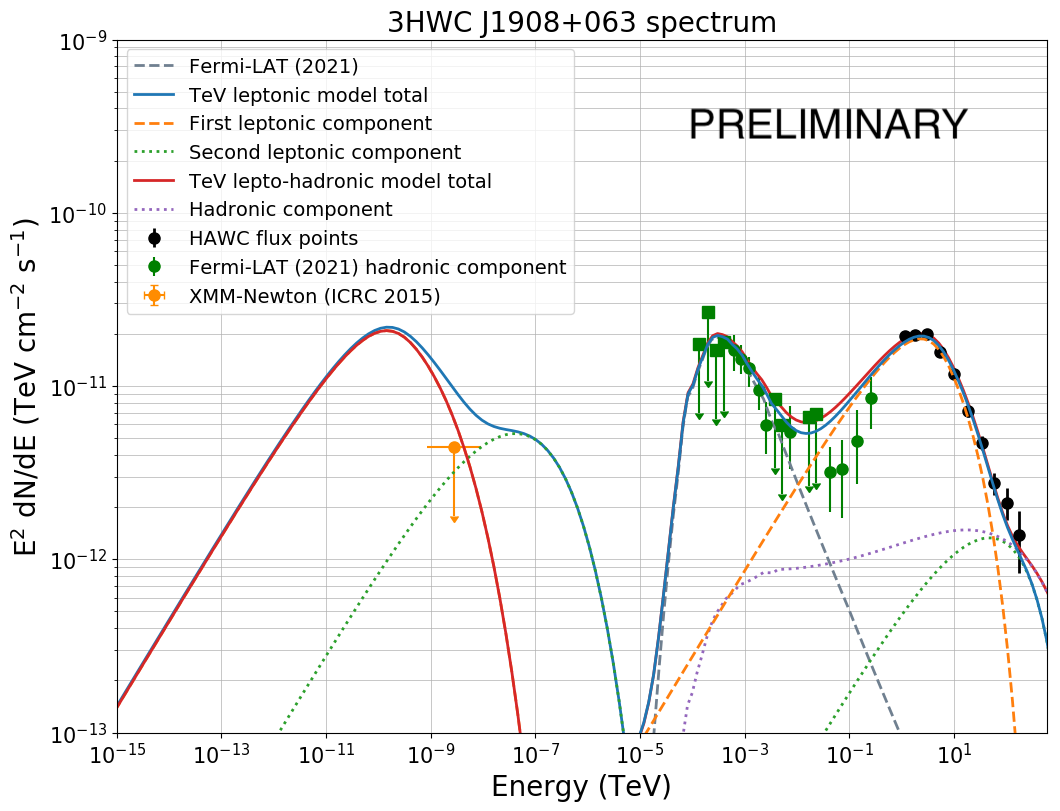}
\caption{Potential models that fit the available multi-wavelength data. The thick red line consists of the hadronic component around 10 GeV from \cite{FermiPWN} along with two leptonic components in the TeV (the sum of the dashed grey + dashed orange + dotted green lines). The thick blue line consists of the hadronic GeV component along with a leptonic component that is prominent in the multi-TeV regime and a second hadronic component at the very highest energies (the sum of the dashed grey + dashed orange + dotted purple lines)}.
\label{fig:comparison}
\end{figure*}

Figure ~\ref{fig:comparison} shows the HAWC spectrum with potential models highlighted. Each model as two components in the TeV range: the thick blue line denotes the two-population leptonic model, while the solid red line denotes a lepto-hadronic model.  Both models contain an additional component in the GeV range, based on recent Fermi data that indicates there is a hadronic component below 10 GeV. Since HAWC is not sensitive to this energy range, we simply use the parameters from ~\cite{FermiPWN}. 

\section{Future outlook}

The uncertainties are presently too large to distinguish whether the component prominent above 50 TeV is leptonic or hadronic in origin. The HAWC Collaboration is currently analyzing a new pass over all collected data, with updated reconstruction algorithms and better gamma/hadron separation algorithms. This may allow for the necessary sensitivity at high energies to distinguish between these two components. HAWC's recently-completed outrigger array will also increase the experiment's sensitivity at the highest energies.

This source is also an attractive target for other experiments. LHAASO has recently detected this source above 100 TeV with a significance of 17.2$\sigma$~\cite{lhaaso2}. Detailed morphology studies above this threshold could help uncover the nature of the source. The two different models presented here start to diverge rapidly just above the energy range that HAWC is sensitive to. 

Multi-wavelength data will also be important. The leptonic model and the lepto-hadronic model have very different fluxes in the keV to MeV energy bands. In-depth measurements from X-ray experiments such as NuSTAR will be useful, as will observations from proposed experiments such as AMEGO~\cite{kierans}. X-ray to GeV observations will also allow for better determination of the spectral shape and energy budget. 

As mentioned in the introduction, this source has long been thought of as a potential neutrino source. Many published predictions of how much data IceCube will need to accumulate before detecting the source assume that it is completely hadronic in nature~\cite{gonzalezgarcia, Halzen2017}.  However, the analysis presented here asserts that the source is dominantly leptonic, with only a fractional hadronic contribution. Those predictions should be re-evaluated in light of this. IceCube-Gen2 may be necessary to detect the source using neutrinos. 

\section{Conclusions}

We report on the spectrum and morphology of 3HWC J1908+063, which is one of the highest-energy known gamma ray sources. It is well-fit to a log-parabola spectrum and a diffusive morphological model. There is evidence that the source exhibits energy-dependent morphology, and there may be hints of spectral hardening at the highest energies. More data, both from HAWC's upgraded array and recently-constructed experiments such as LHAASO, are needed to confirm this. 

It is likely that the source emission originates from mutliple populations of particles, with the bulk of the emission being leptonic and powered by the high-$\dot{E}$ pulsar PSR J1907+0602. However, a hadronic contribution at the highest energies cannot be ruled out. 

\acknowledgments

We acknowledge the support from: the US National Science Foundation (NSF); the US Department of Energy Office of High-Energy Physics; the Laboratory Directed Research and Development (LDRD) program of Los Alamos National Laboratory; Consejo Nacional de Ciencia y Tecnolog\'ia (CONACyT), M\'exico, grants 271051, 232656, 260378, 179588, 254964, 258865, 243290, 132197, A1-S-46288, A1-S-22784, c\'atedras 873, 1563, 341, 323, Red HAWC, M\'exico; DGAPA-UNAM grants IG101320, IN111716-3, IN111419, IA102019, IN110621, IN110521; VIEP-BUAP; PIFI 2012, 2013, PROFOCIE 2014, 2015; the University of Wisconsin Alumni Research Foundation; the Institute of Geophysics, Planetary Physics, and Signatures at Los Alamos National Laboratory; Polish Science Centre grant, DEC-2017/27/B/ST9/02272; Coordinaci\'on de la Investigaci\'on Cient\'ifica de la Universidad Michoacana; Royal Society - Newton Advanced Fellowship 180385; Generalitat Valenciana, grant CIDEGENT/2018/034; Chulalongkorn University’s CUniverse (CUAASC) grant; Coordinaci\'on General Acad\'emica e Innovaci\'on (CGAI-UdeG), PRODEP-SEP UDG-CA-499; Institute of Cosmic Ray Research (ICRR), University of Tokyo, H.F. acknowledges support by NASA under award number 80GSFC21M0002. We also acknowledge the significant contributions over many years of Stefan Westerhoff, Gaurang Yodh and Arnulfo Zepeda Dominguez, all deceased members of the HAWC collaboration. Thanks to Scott Delay, Luciano D\'iaz and Eduardo Murrieta for technical support.



\bibliographystyle{JHEP}
\bibliography{bib}

\clearpage
\section*{Full Authors List: \Coll\ Collaboration}

\scriptsize
\noindent
A.U. Abeysekara$^{48}$,
A. Albert$^{21}$,
R. Alfaro$^{14}$,
C. Alvarez$^{41}$,
J.D. Álvarez$^{40}$,
J.R. Angeles Camacho$^{14}$,
J.C. Arteaga-Velázquez$^{40}$,
K. P. Arunbabu$^{17}$,
D. Avila Rojas$^{14}$,
H.A. Ayala Solares$^{28}$
% ...

\noindent
% ...
$^{14}$Instituto de Física, Universidad Nacional Autónoma de México, Ciudad de México, México,
% ...
$^{17}$Instituto de Geofísica, Universidad Nacional Autónoma de México, Ciudad de México, México,
% ...
$^{21}$Physics Division, Los Alamos National Laboratory, Los Alamos, NM, USA,
% ...
$^{28}$Department of Physics, Pennsylvania State University, University Park, PA, USA,
% ...
$^{40}$Instituto de Física y Matemáticas, Universidad Michoacana de San Nicolás de Hidalgo, Morelia, Michoacán, México,
$^{41}$FCFM-MCTP, Universidad Autónoma de Chiapas, Tuxtla Gutiérrez, Chiapas, México,
% ...
$^{48}$Department of Physics and Astronomy, University of Utah, Salt Lake City, UT, USA,
% ...

\end{document}